\documentclass[pra,showpacs,twocolumn,superscriptaddress,amsmath,amssymb]{revtex4-1}
 
 \pdfoutput=1

\usepackage{graphicx}
\usepackage{wrapfig}
\usepackage{sidecap}
\usepackage{color}
\usepackage{dcolumn,amsmath}
\usepackage{xfrac}

\usepackage{dcolumn}
\usepackage{bm}
\usepackage{amssymb}

\usepackage[caption=false]{subfig}
\usepackage[flushleft]{threeparttable}
\usepackage{array}
\usepackage{adjustbox}

\begin{document}

\title{Prospects for laser cooling of polyatomic molecules with increasing complexity}

\author{Jacek K{\l}os}
\affiliation{Department of Physics, Temple University, Philadelphia, Pennsylvania 19122, USA}
\affiliation{Department of Chemistry and Biochemistry, University of Maryland, College Park, Maryland 20742, USA}
\author{Svetlana Kotochigova}
\email{skotoch@temple.edu}
\affiliation{Department of Physics, Temple University, Philadelphia, Pennsylvania 19122, USA}

\date{\today}

\begin{abstract}
Optical cycling transitions and direct laser cooling have recently been demonstrated for a number of alkaline-earth dimers and  trimer molecules. This is made possible by diagonal Franck-Condon factors between the vibrational modes of the optical transition. Achieving a similar degree of cooling to micro-Kelvin equivalent kinetic energy for larger polyatomic molecules, however, remains challenging. Since polyatomic molecules are characterized by multiple degrees of freedom and have a correspondingly more complex structure, it is far from obvious whether there exist polyatomic molecules that can repeatedly scatter photons. Here, we propose  chemical substitution approaches to engineer large polyatomic molecules with optical cycling centers (OCCs) containing alkaline-earth oxide dimers or acetylenic alkaline-earth trimers (i.e. M--C$\equiv$C) connected to CH$_n$ chains or fullerenes. 
To validate the OCC-character of the selected molecules we performed electronic structure calculations of the equilibrium configuration of both  ground and excited  potential energy surfaces, evaluated their vibrational and bending modes, and corresponding Franck-Condon factors for all but the most complex molecules. For fullerenes, we have shown that  OCCs based on M--C$\equiv$C can perform better than those based on alkaline-earth oxide dimers.
 In addition,  for heavier polyatomic molecules it might be advantageous to attach two OCCs, thereby potentially doubling the photon scattering rate and thus speeding up cooling rates. 
\end{abstract}

\maketitle

\section{Introduction}

Laser cooling in atomic and molecular physics  is the term used to indicate methods to slow and even trap gaseuous atoms with
the forces  generated by the absorption and emission of laser photons, typically in the visible part of the radiation spectrum \cite{Wieman1999}.  Nowadays, it is routinely used to cool and trap alkali-metal, alkaline-earth, noble-gas, and most-recently the heavy magnetic lanthanide atoms \cite{BLev2010, Tm_MOT2010, DyBEC, ErBEC, Lev2012, Ferlaino2012A, Ho_Saffman2014}. Temperatures well below one milli-kelvin at number densities between $10^9$ and $10^{15}$ atoms per cubic centimeter are reached~\cite{Stellmer2013,Urvoy2019}.
Select other atomic species in the periodic table, notably chromium \cite{Griesmaler2005}, have also been laser cooled.
Research with these ultra-cold atomic samples has led to the simulation of exotic phases in quantum-degenerate  gases \cite{Bloch2004}, a novel generation of optical atomic clocks \cite{Ludlow2015}, atom sources for nano-fabrication \cite{Gardner2019}, matter-wave interferometry \cite{PBerman}, and the development of other sensitive detectors \cite{Eckel2018}. 
 
Extremely low atomic temperatures down to a few nanokelvin  combined with the presence of magnetic Fano-Feshbach resonances
\cite{Chin2010,Kotochigova2014} have allowed the efficient creation of samples of di-atomic alkali-metal molecules, by binding  two of these atoms together \cite{Ni2008}. Recently, direct Doppler cooling (the oldest of all means of light-based cooling \cite{Doppler1985}) of  SrF, CaF, YbF, and YO dimers was  demonstrated \cite{DeMille2016,Doyle2,Hinds2017,Collopy2018,Lim2018}. This success is associated with diagonal Franck-Condon factors (FCFs) \cite{Herzberg} on an allowed optical transition between two electronic potentials. Diagonal FCFs ensure that vibrational state $v$ of an excited diatomic electronic potential decays  by  spontaneous emission to the vibrational state ${v'=v}$  of the ground electronic potential with near unit probability, once the rotational levels of the ground and excited electronic states have been carefully chosen based on the selection rules of the electric dipole transition. The nearly-closed two-level system can absorb and emit thousands of photons per molecule to achieve cooling. The remaining spontaneous decay populates unwanted states such as other vibrational and hyperfine states of  energetically-lower potentials.

Over the past decades spectroscopic studies of polyatomic molecules consisting of an alkaline-earth metal (M) and a ligand (L) have attracted particular attention after the discovery of their peculiar ionic chemical bond \cite{PRESUNKA1995,WORMSBECHER1,WORMSBECHER2,HILBORN,BERNATH1984,Brazier1987,Bopegedera1987,Brazier1989,BERNATH1991}. High-resolution laser-spectroscopy experiments on CaOH (and CaOD)  \cite{HILBORN} and SrOH (and SrOD) \cite{NAKAGAWA1983} were the first to deduce the strong ionic character of metal-hydroxide bonds. More recently, this led to  suggestions of efficient optical cycling and laser cooling of these polyatomic molecules published in two conference proceedings \cite{Isaev2014,Baum2015}. The authors of Ref.~\cite{Kozyryev2015} measured FCFs near unity for transitions in the linear SrOH molecule and suggested that this molecule should be able to scatter thousands of photons before transitioning into unwanted vibrational states. The first optical cycling and radiation pressure for polyatomic molecules (SrOH) was  introduced in Ref.~\cite{Kozyryev2016b}. Theoretically, this idea was further investigated in Refs.~\cite{Kozyryev2016,Isaev2016}.  Within a year optical cycling in  SrOH  was observed  by Doyle's group \cite{Doyle2017}. In this experiment the transverse temperature of a cryogenic SrOH molecular beam was reduced in one dimension by two orders of magnitude to 750 $\mu$K by laser cooling.  Recently, the same group  succeeded in laser cooling of  polyatomic YbOH  to temperature below 600 $\mu$K \cite{Augenbraun2020}. This molecule is expected to have an exceptionally high sensitivity to physics beyond the Standard Model \cite{Kozyryev2017}. For such promise of sensitivity to be observable large numbers of YbOH are required.  In this context it is worth nothing that Ref.~\cite{Jadbabaie2020} developed a method that improves the yield of YbOH by using laser light to excite a metal atom precursor. 

Few polyatomic systems  actually have cycling transitions, now called optical cycling centers (OCC) as the relevant transition is localized at a single bond 
or even on a single atom within the molecule. In this paper, OCCs describe a compact quantum functional object, which can repeatedly experience laser excitation and spontaneous decay, while being well isolated from its environment.  They can be attached to either large ligand-molecules or to even ligand-surfaces allowing trapping and cooling of chemically diverse species. So far, only (M)-monohydroxides and  larger (M)-monoalkoxide molecules with  metallic atom M given by Ca, Sr , Ba, or Yb have been suggested as good candidates \cite{Kozyryev2016,Kozyryev2017}. It is now highly desirable to identify other classes of polyatomic molecules with OCCs. 
The essential property of polyatomic molecules with OCCs is the nonbonding s-electron of the alkaline-earth (AE) or rare-earth (RE) atom within the molecule being removed from the bonding region by orbital hybridization \cite{Ellis2001}, leading to highly diagonal FCFs. This property is not strongly dependent on the type of functional group bound to the metal atom \cite{Isaev2016,Kozyryev2016}. Our recent research on electronic, vibrational, and rotational structure of both ground and excited states of the SrOH molecule \cite{Klos2019} allowed us to gain a quantitative understanding of the localized OCC near its Sr atom. 

In this paper we show that two approaches to engineer polyatomic molecules of increasing complexity with usable OCCs exist.
This engineering approach may have wider technological applications \cite{Toworfe2009,Lessel2015}.
First, following \cite{Kozyryev2016,Kozyryev2019} we replace the hydrogen atom in Ca-O-H by a larger ligand, molecules or chains of molecules, that, as we will show, do not significantly disturb the valence electron of the metal-cation Ca-O bond so that optical cooling  remains possible. In particular, we examine  the increasing complexity of the electronic structure and its effect on the optical cycling transitions in the family of molecules Ca-O-(CH$_2$)$_n$-CH$_3$, where ${n= 1}$ or 2. Second, we describe a next generation of candidate molecules, one with two OCCs attached to the ligand, with the goal to double the cooling force. Finally, we investigate the attachment of OCCs to exohedral and endohedral fullerene molecules. We  compare results for two OCCs, i.e. the strontium acetylenic carbon chain Sr-C$\equiv$C and  strontium oxide Sr-O.

\section{Optical cycling centers in the C\lowercase{a}-O-(CH$_{\bf 2}$)$_{\bf n}$-CH$_{\bf 3}$ molecular family}

In this section we study the optical cycling transition and other molecular properties of the family of polyatomic molecules Ca-O-(CH$_2$)$_n$-CH$_3$, where 
${n=1}$ or 2.  These molecules are characterized by the replacement of the hydrogen atom in CaOH by a larger ligand \cite{Bernath1986}.  We performed electronic structure calculation of these molecules using density functional theory (DFT) within the Gaussian-09 program \cite{Gaussian09_RevE}. 
The correlation-consistent polarized valence triple-zeta (cc-pVTZ) basis for the Ca atom and Dunning's augmented with additional diffuse functions (aug-cc-PVTZ) basis for the H, C and O atoms have been used. The calculations include geometry optimization of ground and excited states using time-independent and time-dependent DFT, respectively.  In both cases the spin-unrestricted WB97XD functional is employed. We  computed  electronic potentials near their equilibrium geometry as well as  permanent and transition electric dipole moments. For example, Table \ref{bond_length} lists the Ca-O bond separations in the optimized structure of the ground X and lowest two excited states, denoted A and B, respectively, of four Ca-O-R molecules. 
We first observe that the bond lengths, regardless the complexity of the molecule or  electronic excitation, agree within $\approx$1\%.
Secondly, the Ca-O bond lengths for the X and B states are consistently larger than those for the A state, and, in fact, the separations 
for the X and B states agree to better than 0.2\%. As we will show this  ultimately leads to more diagonal FCFs for the X-B transition than for the X-A transition. The bond lengths for CaO-CH$_3$ are always smallest. 

\begin{figure*}
\includegraphics[width=1.2\columnwidth,trim=0 25 20 40,clip]{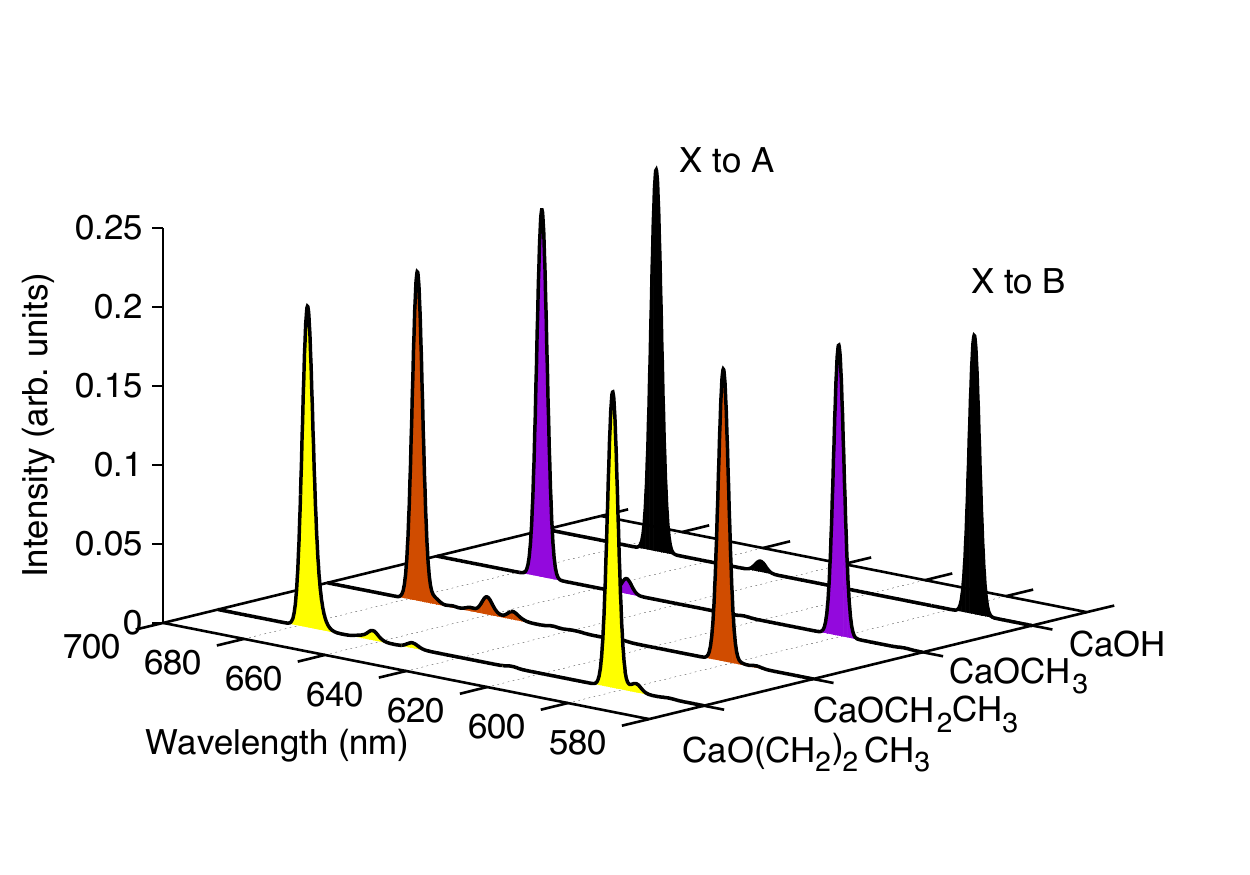}
\includegraphics[width=0.7\columnwidth,trim=15 0 0 0,clip]{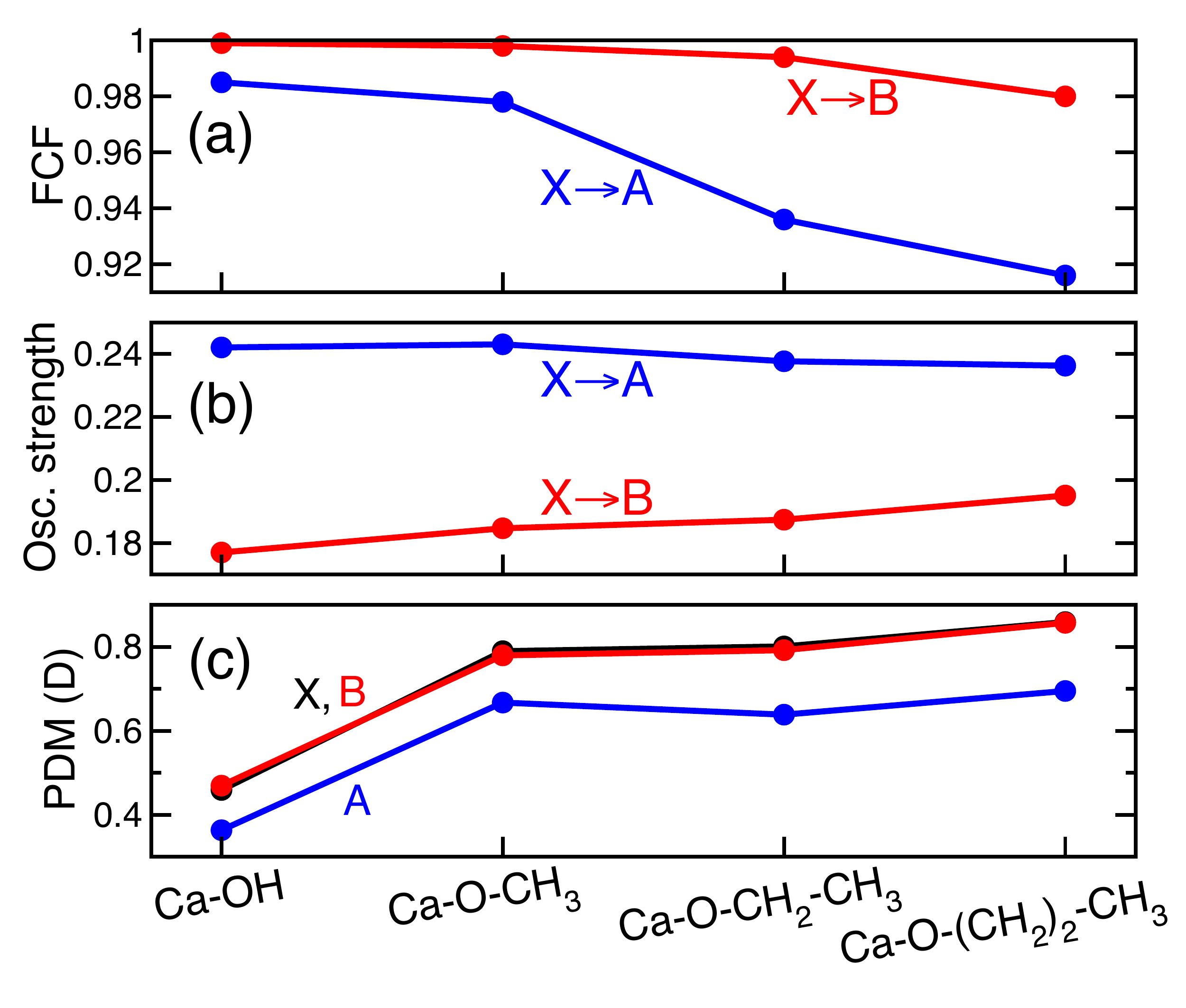}
\caption{Left panel: Simulation of line intensities as functions of laser wavelength from the X(0,0,0) ground vibrational state to vibrational levels of the A(0,0,0) and B(0,0,0) electronic states for CaOH, CaO-CH$_3$, CaO-CH$_2$-CH$_3$, and CaO-(CH$_2$)$_2$-CH$_3$.  We  assume that each transition contributes a Gaussian with height equal to its line strength and a full width at half maximum (FWHM) width of 3 nm, typical of  widths observed in Ref.~\cite{Kozyryev2019}.
 Right panel: Franck-Condon factors (panel (a)), oscillator strengths (b), and permanent dipole moments in Debye (c) for the relevant transitions in CaOH, CaO-CH$_3$, CaO-CH$_2$-CH$_3$, and CaO-CH$_2$-CH$_2$-CH$_3$.  In panels (a) and (b)  blue and red curves correspond to the cycling transition from the X(0,0,0) to A(0,0,0) and X(0,0,0) to B(0,0,0)  vibrational levels, respectively. In panel (c) the black, blue, and red curves are permanent dipole moments for the X(0,0,0), A(0,0,0), and B(0,0,0)  levels, respectively.}
\label{fig:ExcitationSpectra}
\end{figure*}

\begin{table}
\caption{Equilibrium separations for the Ca-O bond in the CaO-R molecular family for the electronically ground X and excited A and B states.}
 \label{bond_length} 
\vspace*{0.2cm} 
\begin{tabular}{|c|c|c|c|}  \hline 
Ca-O separation  & X         &  A       &    B  \\ 
($10^{-10}$~m) &&&\\
 \hline 
CaOH                     & ~1.974~   & ~1.958~  & ~1.976~ \\ [1 pt]
\hline
CaO-CH$_3$          & 1.972   & 1.953  &  1.968 \\ [1 pt]
\hline
CaO-CH$_2$CH$_3$ & 1.974 & 1.954 & 1.972 \\ [1 pt]
\hline
CaO-(CH$_2$)$_2$CH$_3$ & 1.973 & 1.953 & 1.974 \\ [1 pt]
 \hline  
 \end{tabular}      
\end{table}

Next, we compare oscillator strengths or intensities \cite{Herzberg} of electronic transitions from the ground X(0,0,0) vibrational normal mode to  vibrational normal modes $(v_1,v_2,v_3)$ of the excited A and B states of CaO-R molecules. By the notation (0,0,0) and $(v_1,v_2,v_3)$, inherited from the CaOH trimer, we describe  the quantum numbers for the normal modes associated with the Ca-O stretch, Ca-O-C bend, and O-C stretch of the ground and excited states, respectively. Here, C is the carbon atom of ligand R directly bound to the CaO fragment.  In the Ca-O-R series the normal mode of the X state with the lowest vibrational energy is associated with the Ca-O-C bend motion. For example, for Ca-O-CH$_3$ the vibrational energy equivalent in the Ca-O-C bend is 147 cm$^{-1}$. The next, larger frequency for Ca-O-CH$_3$ is 475 cm$^{-1}$ and associated with the Ca-O stretching mode. The O-C stretching mode is much higher in energy, with energy equivalent 1187 cm$^{-1}$. The normal mode energies remain similar with increasing size of R, but tend to decrease in energy.  

We find that most of transitions between ground- and excited-state normal modes that have significant Franck-Condon factors are associated with $(v_1,v_2,v_3)=(0,0,0)$ mode and to a lesser degree for $v_1>0$ levels in  the Ca-O stretch of the A or B excited state. Transitions to  O-C stretch modes with $v_3>0$ have  Franck-Condon factors on the order of 0.05 out of 1.0. Franck-Condon factors for transitions from the ground X(0,0,0) state to the A or B normal modes associated with internal vibrational motion of R are negligibly small (e.g. $<$ 0.0001). The same goes for electronic transitions involving excitations within the R fragments, they are higher in energy than electronic excitations localized on the OCC. The effect of a change in rotational state of the molecules on the transition dipole moment can simply be computed with angular momentum algebra.

The left panel of Figure~\ref{fig:ExcitationSpectra} shows simulated spectra in the visible wavelength domain near the optical-cycling transitions for CaOH and increasing molecular complexity of Ca-O-(CH$_2$)$_n$-CH$_3$. The molecules have a single optical center. The excitation spectra are obtained from electronic transition oscillator strengths in the double-harmonic approximation using the eZspectrum program \cite{Mozhayskiy}. The stick-like spectra have been convoluted with a unit-normalized Gaussian function to obtain smooth curves. Several observations can be made. The first is that transition frequencies and oscillator strengths for the X-A and X-B transitions only slightly change along the CaO-R series. In other words, the complexity of the ligand does not significantly affect the transition. The shift in transition frequency, however, is large compared to their natural linewidth and can be readily resolved. Secondly, one line for the X-A and X-B transitions dominates with smaller features appearing at smaller wavelengths.  The peaks of smaller intensity  between 650 nm and 660 nm that appear between the two largest peaks shown in the left panel of Fig.~\ref{fig:ExcitationSpectra}  originate from transitions from the X(0,0,0) to normal modes of the electronic-exited states associated with the Ca-O stretch (A(1,0,0) or for longer  R chains with Ca-O stretch and bend  modes, such as A(1,1,0)).

The small number of strong lines in Fig.~\ref{fig:ExcitationSpectra} reflect the diagonal nature of the Franck-Condon factor. These observations are made quantitative in the right panel of Fig.~\ref{fig:ExcitationSpectra}. The figure shows our values for the Franck-Condon factors, oscillator strengths, and permanent dipole moments between vibrational modes X(0,0,0), A(0,0,0) and B(0,0,0). The Franck-Condon factors are close to one but decreasing with complexity indicating that longer chains of ligands do change the coupling between Ca and O and make cooling less efficient. The B state is more favorable for laser cooling. The FCF values for the X to B state are nearly the same for all four molecules. The permanent dipole moment can be used to calculate the infra-red transition strength between vibrational states of the same electronic potential. It also is a measure of the electron distribution in the bond between Ca and O. In principle, the X-B transition can be followed by spontaneous emission to the A state. This can reduce the efficiency of the cooling process in addition to the efficiency reduction due to the breakdown of diagonality between the X(0,0,0) and B(0,0,0) states. A simple estimate, however, shows that the rate of such transitions is $10^3$ times smaller than those for decay to the X state. Spontaneous emission rates are proportional to $\omega^3d^2$, where $\omega$ is the transition frequency and $d$ is the transition dipole moment. The dipole moments for the B to A and B to X transitions are similar in magnitude.  The transition frequency for the B to A transition, however, is ten times smaller than that for the B to X transition.

Finally, we have compared our theoretical values of FCFs for the CaOH and CaOCH$_3$ molecules shown in Fig.~\ref{fig:ExcitationSpectra}b with recent experimental measurements of FCFs for the X-A transition in Ref.~\cite{Kozyryev2019}. The agreement between the numbers is  3\% and 5\%, respectively. This is quite satisfactorily, but a possible explanation for the disagreement is the absence of spin-orbit interactions in our formalism.

\section{Two optical cycling centers in the C\lowercase{a}-O-(CH$_{\bf 2}$)$_{\bf n}$-CH$_{\bf 3}$-O-C\lowercase{a}(S\lowercase{r}) molecular families}

For heavier molecules it can be advantageous to attach two OCCs, thereby possibly doubling the photon scattering rate. 
The two optical centers M and M$^\prime$ are expected to cycle photons almost independently as the bond to their oxygen atom are (mostly) independent of each other. This independence is more likely true when M and M$^\prime$ are not the same species. We can then use two lasers with distinct colors and, thereby, increase the cooling 
force on multi-center molecules. When M and M$^\prime$ are the same atom and the symmetry of the molecule is sufficiently high, we must consider hybridization of the excited states as we cannot distinguish which center is excited. This hybridization might be describable in terms of superradiance and bright and dark states. These assertions for large molecules with one-, two-, or even more cooling centers are all unverified and conjecture at this stage. During the research period we hope to supply answers to some of these questions. 
The hybridization in symmetric two-center molecules that leads to the doubling in line intensity can be explained as follows. Each cycling center has a ground- and excited-electronic state, $|g\rangle$ and $|e\rangle$, respectively. Two nearly-degenerate excited molecular states, conveniently labeled by  {\it gerade} ({\it g}) and {\it ungerade} ({\it u}), exist. They are written as $|\phi^{(e)}_{g,u}\rangle=( |eg \rangle \pm |ge\rangle )/\sqrt{2}$ in the two separate-emitter basis.  If there is no coupling between the emitters  $|\phi^{(e)}_{g,u}\rangle$, then the two states  are degenerate. With non-zero coupling, which is the case for our molecules, the two levels are split. The ground-state wave function in the same separate-emitter basis is 
$|\phi^{(g)}\rangle = |gg\rangle$. Then the electronic transition dipole moment  $\langle \phi^{(g)} | d | \phi^{(e)}_{g,u} \rangle$ for the bright {\it gerade} state is $\sqrt{2} \langle g|d|e\rangle$ and zero for the dark {\it ungerade}  state. Consequently, the transition dipole moment is amplified to 1.4 times that of a single emitter, and the Rabi frequency is two times larger. The validity of this description relies on symmetry within the molecule as well as the separation between the optical centers. With a small ligand the two optical centers might be relatively close and share a dipole moment.

\begin{figure}
 \includegraphics[width=0.97\columnwidth,trim=5 15 0 0,clip]{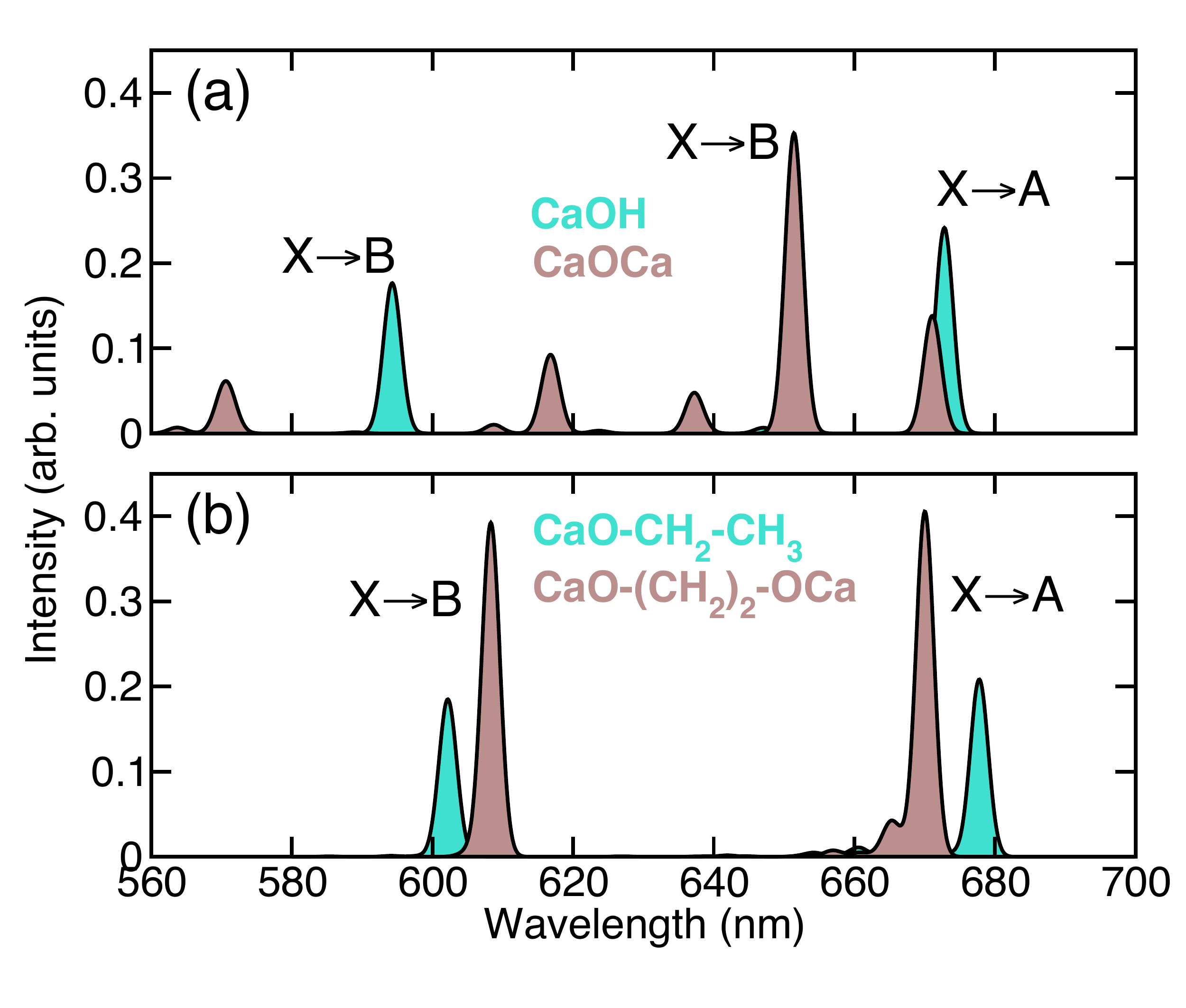}
 \includegraphics[width=0.93\columnwidth,trim=0 0 0 15,clip]{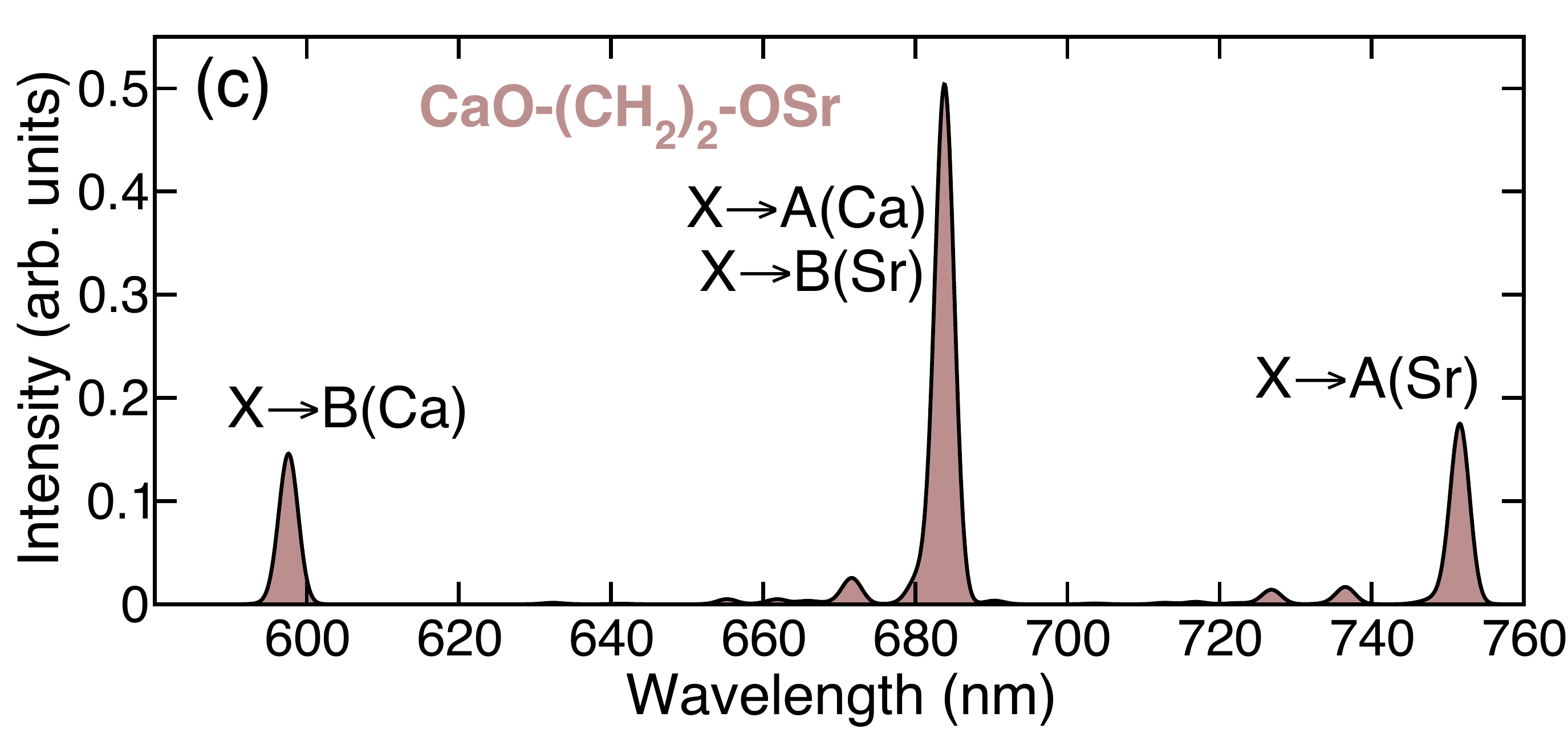}  
 \caption{Theoretical excitation spectra for molecules with one and two optical-cycling centers as functions of laser wavelength assuming Gaussian line profiles with a FWHM width of 3 nm as in Fig.~\ref{fig:ExcitationSpectra}. Panel (a) compares the spectra for CaOH with that for CaOCa. Panel (b) compares Ca-O-CH$_2$-CH$_3$ with Ca-O-(CH$_2$)$_2$-O-Ca. Panel (c) shows the excitation spectrum for the asymmetric molecule Ca-O-(CH$_2$)$_2$-O-Sr. In all panels transitions from X(0,0,0) to A(0,0,0)  and B(0,0,0) are strongest. In panel (c) the X to A(Ca) and A to B(Sr) transitions, where the Ca and Sr atom is excited, respectively, are degenerate within the width of the Gaussian line shape. }
 \label{fig:TwoCenters}
\end{figure}

\begin{table*}
\caption{Oscillator strengths and Franck-Condon factors (FCF) for the fundamental X(0,0,0)-A(0,0,0) and X(0,0,0)-B(0,0,0) transitions in two optical-center compounds CaOCa and MO-R-OM$^\prime$. For 
CaO-(CH$_2$)$_2$-OSr columns labeled Sr and Ca indicate the atom that is predominantly excited.}
\label{tab:FCFAB}
\begin{center}
\begin{tabular}{|c|c|c|c|c|c|c|c|c|} \hline 
 OCC complex  & \multicolumn{4}{c|} {X - A }           & \multicolumn{4}{c|}{X - B}\\[1 pt]
 \hline 
                                   &   \multicolumn{2}{c|}{ Osc. strength}      & \multicolumn{2}{c|}{FCF}        &  \multicolumn{2}{c|}{Osc. strength}   &  \multicolumn{2}{c|}{FCF}\\[1pt]
                                    \hline
 CaOCa           &   \multicolumn{2}{c|}{0.493 }               &  \multicolumn{2}{c|}{0.530}     &  \multicolumn{2}{c|}{0.403}           &  \multicolumn{2}{c|}{0.936} \\[1 pt]
 \hline
 CaO-(CH$_2$)$_2$-OCa &\multicolumn{2}{c|}{0.468} &\multicolumn{2}{c|}{0.93} & \multicolumn{2}{c|}{0.400} & \multicolumn{2}{c|}{0.990}\\[1 pt]
  \hline                                              
                                       & Sr          &  Ca       &  Sr      & Ca  & Sr         &  Ca         &  Sr      & Ca\\[1 pt]
                                        \hline 
 CaO-(CH$_2$)$_2$-OSr  & 0.226         & 0.296         &   0.879     & 0.919   & 0.283        &  0.149         &  0.929     &  0.992 \\
 \hline          
\end{tabular}
\end{center}
\end{table*}

Here, we focus on the family  Ca-O-Ca, Ca-O-CH$_2$-O-Ca, and Ca-O-CH$_2$-CH$_2$-O-Ca molecules with two identical Ca atoms or two different atoms (Ca and Sr) as the OCCs.
In Table~\ref{tab:FCFAB} we list oscillator strengths and FCF values for the fundamental vibrational transitions in X(0,0,0)-A(0,0,0) and X(0,0,0)-B(0,0,0) electronic excitations for three molecules MO-R-OM$^\prime$ with two OCCs, where M is not necessarily equal to M$^\prime$. 
The data quantifies the observations in Fig.~\ref{fig:TwoCenters}. This figure shows that cycling rates can indeed be increased with two optical centers.   In panel (a) we compare the theoretical spectrum for the symmetric, Ca-O-Ca, where its cycling centers are close together, with that of Ca-OH. The X(0,0,0) to A(0,0,0) transition near 670 nm in the Ca-O-Ca spectrum has weakened and several relatively strong features at shorter wavelengths are visible. They are due to bending modes in the A state. On the other hand, the X(0,0,0) to B(0,0,0) transition near 650 nm has remained nearly diagonal but has shifted to significantly longer wavelengths. In fact, its line intensity has doubled compared to the same transition in CaOH. The transition can be thus used as a cycling center. Panel (b) compares theoretical spectra for Ca-O-CH$_3$ and CaO-CH$_2$-OCa. The two Ca atoms in the latter molecule are much further apart than in Ca-O-Ca. In this case the two spectra are far more similar and, crucially, for both the X to A and X to B transitions have doubled in strength. The FCF are also more diagonal.  

In summary, the oscillator strength of the symmetric Ca-O-Ca and Ca-O-(CH$_2$)$_2$-O-Ca are twice that of the single-center Ca-O-R molecule.  In case of the heteronuclear Ca-O-(CH$_2$)$_2$-O-Sr with a corresponding simulated spectrum  that is shown in panel (c) of Fig.~\ref{fig:TwoCenters}, there is no factor of two enhancement in the oscillator strengths as the broken symmetry lifts the degeneracy.

\section{Exohedral  and endohedral fullerenes with OCCs}

In this section, we investigate the possibility of attaching OCCs to one or two of the carbon atoms in an exohedral or endohedral fullerene molecule C$_{60}$ using chemical 
engineering approaches \cite{Meyer2002}. The attraction of fullerenes lies in their relatively-simple electronic structure and high symmetry, 
which might enable multiple applications \cite{Klos2016,Liu2017,Popov2017}. For example, endohedral fullerenes, those that encase a suitable atom,  are  
candidates for building a quantum computer \cite{Benjamin2006,Jones2006,Brown2010,Schoenfeld2006,Farrington2012,Mitrikas2014,Popov2017}. Reference~\cite{Benjamin2006} has suggested that chains of endohedral fullerenes are suitable for storing and manipulating quantum information. 

Here,  we computationally model a fullerene with or without a caged N($^4$S) atom and with or without attached acetylenic C$\equiv$C-Sr or metal-oxide O-Sr groups. These 
groups have an OCC on the Sr atom. We study the cases of either a single or a double group. The ground states of endohedral N@C$_{60}$  and  N@C$_{60}$-C$\equiv$C-Sr complexes are  spin quartet and sextet states, respectively, while both C$_{60}$--C$\equiv$C-Sr and Sr-C$\equiv$C--C$_{60}$--C$\equiv$C-Sr are spin triplets. The fully spin polarized quintet for the double-OCC complex lies $E/hc=3.3$ cm$^{-1}$ above the triplet state.  

We  perform geometry optimization  and frequency calculations of normal modes for the electronic ground state of these molecular systems using the spin-unrestricted 
Coulomb-attenuated UCAM-B3LYP density functional within the DFT approach. The atomic basis set  6-31G(d) is used for the non-metallic atoms while for 
Sr we use the effective core potential ECP28MDF \cite{Kaupp1991} along with the (6s6p5d)/[4s4p2d] PP basis. 

\begin{figure*}
 \includegraphics[scale=0.21]{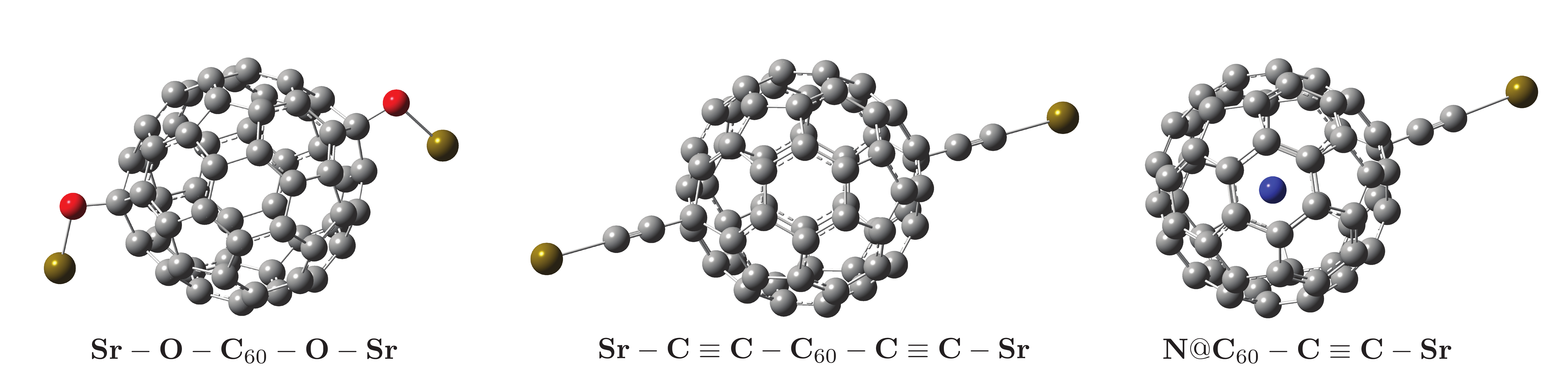}
 \caption{Optimized geometries of  the SrO and Sr-C$\equiv$C OCC derivatives of  C$_{60}$ and N@C$_{60}$. }
 \label{fig:C60geometries}
\end{figure*}

Figure~\ref{fig:C60geometries} shows the optimized geometries of  Sr-O and Sr-C$\equiv$C- OCC derivatives of C$_{60}$ and N@C$_{60}$.  We find that the metal oxide OCC connected to  C$_{60}$ is not optimal as  the Sr  atom  bends towards the fullerene carbon shell, which has unfavorably effects on the metal cycling center. On the other hand, the acetylenic C$\equiv$C linker shown in the figure behaves more favorably by being rigid and keeping the Sr cycling center away from the carbon shell preventing extensive mixing of Sr and C$_{60}$ electron orbitals.

\begin{figure}
 \includegraphics[width=0.97\columnwidth,trim=5 0 0 0]{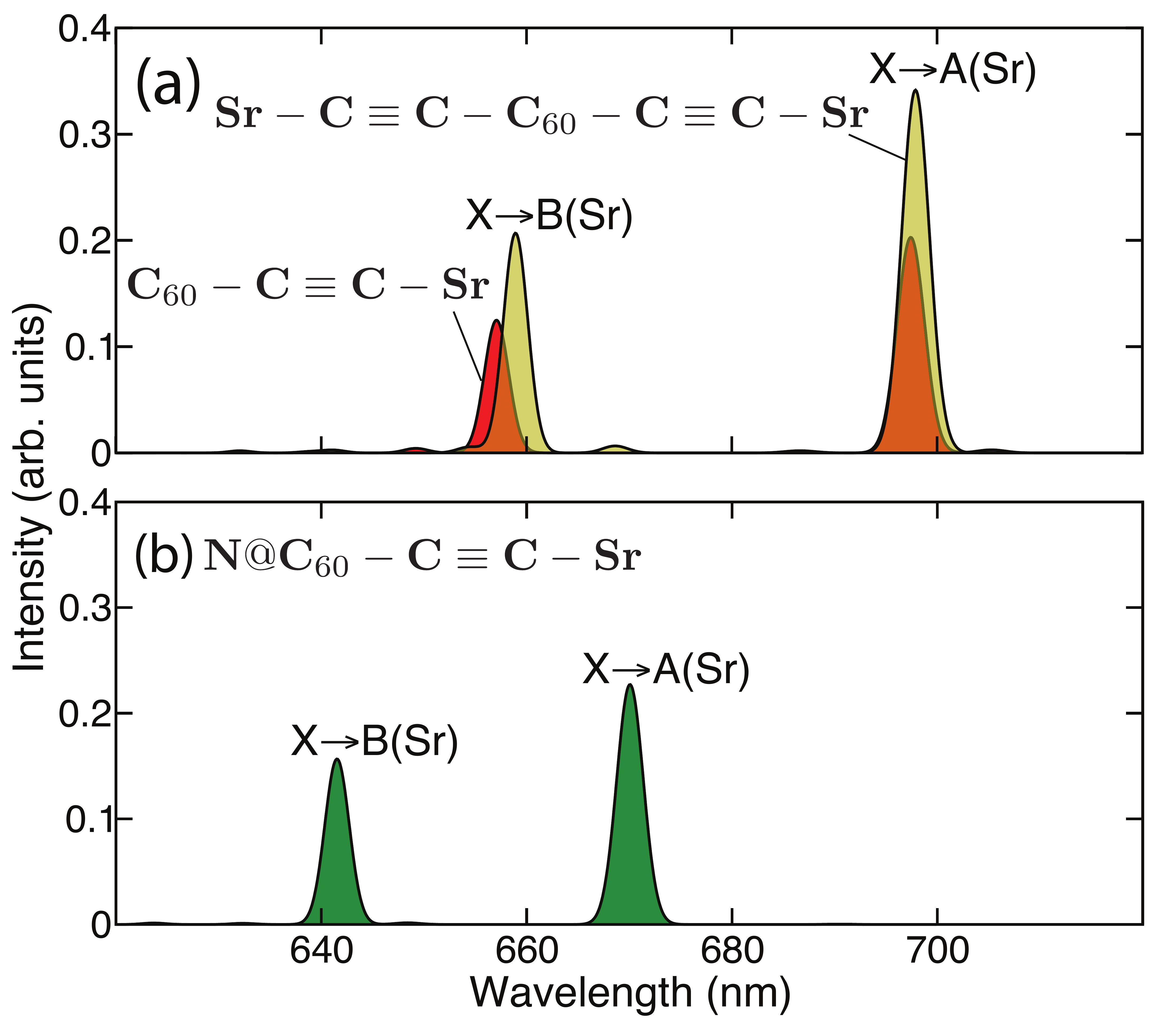}
 \caption{ Panel (a): Optical absorption spectra for  spin triplet  C$_{60}$--C$\equiv$C-Sr (orange) and Sr-C$\equiv$C--C$_{60}$--C$\equiv$C-Sr (yellow),  single  and double OCC  derivatives of C$_{60}$, respectively. Panel (b): Absorption spectrum for the spin sextet OCC C$\equiv$C-Sr derivative of  N@C$_{60}$. In both panels  a 3-nm FWHM broadening of the stick spectrum of calculated oscillator strengths is applied.}
 \label{fig:C60OCCs}
\end{figure}

Electronic excited states and optical spectra in the visible domain of OCC-containing fullerene complexes at the optimized geometry of the electronic ground 
state have been computed using time-dependent DFT. 
For all complexes the lowest 30 roots have been selected for the diagonalization of the electronic Hamiltonian. The calculations provide excitation 
energies along with oscillator strengths for electric-dipole-allowed transitions between ground and excited states. In Table~\ref{tab:OCCC60} we list permanent dipole moments for the C$_{60}$-OCC compounds along with the oscillator strengths for the major electronic transitions visible in the spectra shown in Fig~\ref{fig:C60OCCs}. The large dipole moments for the C$_{60}$-OCC complexes are due to the larger separation between center of mass and center of charge in these  complexes.
The calculated spectra, constructed from these transition energies and oscillator strengths, for single- and double-OCC derivatives of the C$_{60}$ are 
shown in Fig.~\ref{fig:C60OCCs}a, while the spectrum for the single-OCC derivative of N@C$_{60}$ is shown in Fig.~\ref{fig:C60OCCs}b. 
For all complexes two transitions are dominant. They correspond to the X$\rightarrow $A and X$\rightarrow$B transitions, respectively.
The line intensities of the  double-OCC derivative are approximately twice as large as that for the single-OCC derivative again due to the effects of 
superradiance. The spectrum for  N@C$_{60}$  is qualitatively similar to the equivalent structure without the nitrogen atom, although the wavelengths 
of the  X$\rightarrow $A and X$\rightarrow$B transitions have shifted towards higher energies. 

\begin{table*}
\caption{Permanent dipole moments ($\mu$) and oscillator strengths for the fundamental X(0,0,0)-A(0,0,0) and X(0,0,0)-B(0,0,0) transitions in one and two optical-center compounds of C$_{60}$.}
\label{tab:OCCC60}
\begin{center}
\begin{tabular}{|c|c|c|c|} \hline 
 OCC-C$_{60}$ complex  &$\mu$ / Debye& \multicolumn{2}{c|}{Osc. Strength}\\[1 pt]
 \hline 
                                  & &   {X - A }         &   {X - B }   \\[1pt]
                                    \hline
 C$_{60}$-C$\equiv$C-Sr    &5.85       &  0.169                 &  0.203           \\[1 pt]
 \hline
 N@C$_{60}$-C$\equiv$C-Sr&6.82&0.166 & 0.227\\[1 pt]
  \hline                                             
                                   
 Sr-C$\equiv$C-C$_{60}$-C$\equiv$C-Sr  &0.00& 0.295                 & 0.422            \\
 \hline          
\end{tabular}
\end{center}
\end{table*}

\begin{figure*}
 \includegraphics[scale=0.16]{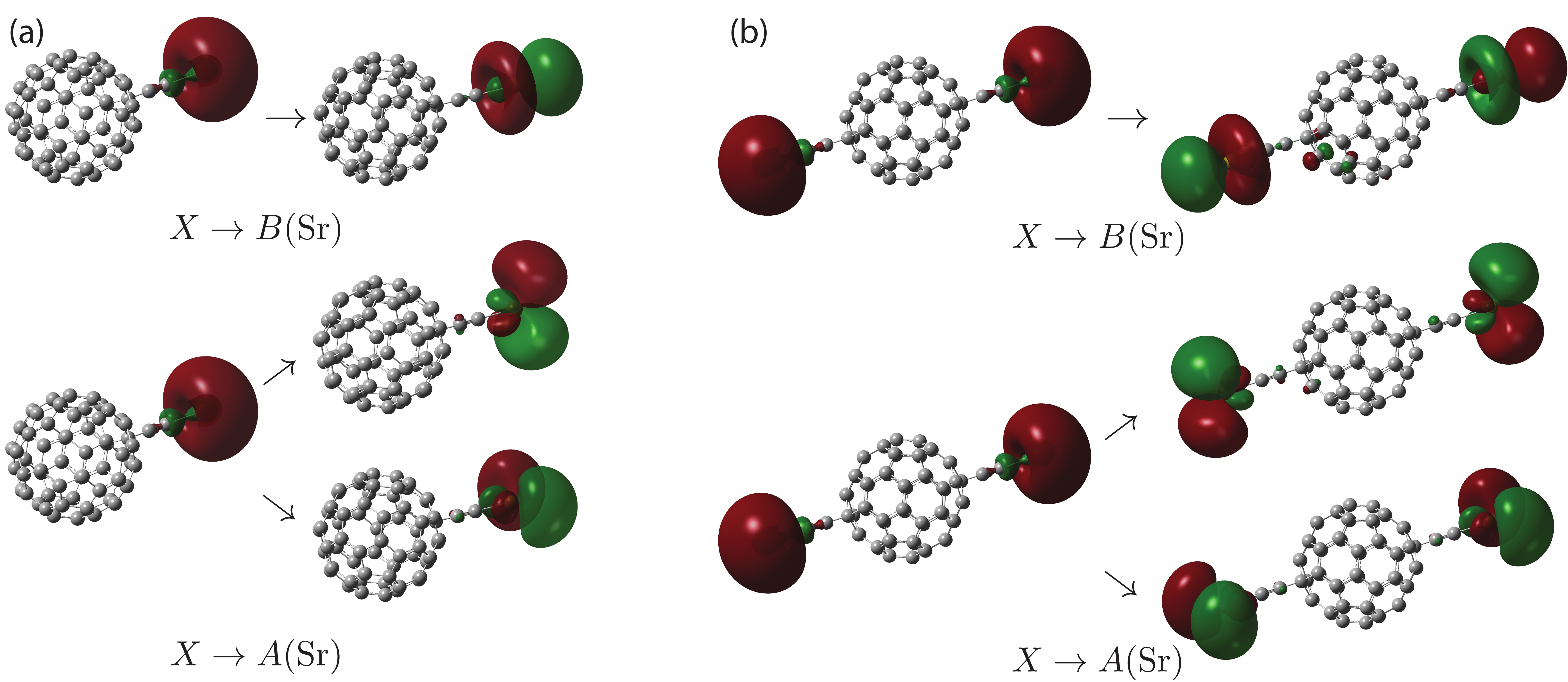}
 \caption{Natural transition orbitals (shown as the red/green isosurfaces at $\pm0.02$, respectively) associated with X$\rightarrow$ A(Sr) and X $\rightarrow$ B(Sr) electronic excitations for (a) the C$_{60}$--C$\equiv$C-Sr single-OCC, (b)  the  Sr-C$\equiv$C--C$_{60}$--C$\equiv$C-Sr double-OCC derivatives. Gray spheres connected by lines represent the location of the 62 or 64 carbon atoms. In all cases the transitions originate from the HOMO orbital mainly described by the 5s  orbital of the Sr atom(s) and end up in
a natural orbital that corresponds to different orientations of a mainly 5p   orbital of Sr.
  For the X$\rightarrow$ A(Sr) transition,  two nearly-degenerate transitions exist.}
 \label{fig:NTOs}
\end{figure*}

To better analyze the laser cooling properties of fullerenes we determined natural transition orbitals (NTOs) of the electronic transitions with significant oscillator strengths. NTOs for optimized geometries of the electronic ground state of C$_{60}$--C$\equiv$C-Sr and 
Sr-C$\equiv$C--C$_{60}$--C$\equiv$C-Sr are shown in panels (a) and (b)  of Fig.~\ref{fig:NTOs}, respectively. The NTOs of the endohedral N@C$_{60}$--C$\equiv$C-Sr are  similar to those shown in panel (a) and not shown  for clarity. In all three cases, the Sr atoms are kept away from the C$_{60}$ by the rigid triple C$\equiv$C alkyne bonds and the complex forms a quasi-linear molecule.  This results in the preservation of the OCC as one can note from the figure.  The NTOs participating in the electronic transition relevant to optical cycling are mostly localized  on the Sr atom. There are tiny contributions, visible on the C atom of the C$\equiv$C linker from the $\pi$ orbitals of the triple bond and for the double OCC case (panel (b)) a tiny contribution in the X$\rightarrow$B(Sr) transition from the C$_{60}$ shell. This property makes the C$\equiv$C-Sr OCC center a promising candidate for the purpose of laser cooling of fullerenes.  In contrast, we find that replacing the triple-bonded C$\equiv$C with a single oxygen atom causes the Sr atom to bend towards the electron-rich C$_{60}$ as already shown in Fig.~\ref{fig:C60geometries}. This behavior  destroys the separation of the metal of OCC from the polyatomic fragment and  destruction of the laser cooling scheme.

Finally, we would like to highlight the challenges associated with laser cooling of large molecules. Increasing the mass of the molecule also increases the number of scattered photons required to reach the same change in temperature or momentum. Hence, non-laser-based cooling techniques will remain an essential ingredient in cooling of heavy molecules. There is also the potential of stimulated optical forces with these large molecules. For example, the bichromatic force has been used to control SrOH \cite{Kozyryev2018} and  CaF \cite{Galica2018}. A related technique, using a mode-locked laser, has  recently been demonstrated for Rb atoms \cite{Long2019}. In addition, it might be possible to attach more than two OCC centers to the fullerene cage. However, the OCCs have to be  spatially well separated in order to avoid direct electronic interactions between them that could destroy their independent laser cycling activity. Also, the linker should be rigid, as in the case of the C$\equiv$C linker, in order to prevent bending of the  OCC.

\section{Conclusion}

We have shown strong evidence that optical cycling can be achieved in quite complex molecules containing metallic alkaline-earth atoms, Ca or Sr, connected to O-CH$_n$ chains or to C$\equiv$C--C$_{60}$ as indicated by the strongly diagonal Franck-Condon factors.  We have determined the degree to which the OCCs of the studied molecules have the requisite diagonal Franck-Condon factors by performing electronic structure calculations near the equilibrium configuration of multi-dimensional  ground and excited potential energy surfaces and  evaluating their vibrational and bending modes. We find that the OCC attached to the polyatomic molecules of increasing complexity and number of normal modes retains its character necessary for the laser cooling. The increase of normal modes doesn't  significantly disturb  the two-level system with diagonal Franck-Condon factors, but there might be some possibility for leaking to the bending normal modes with increasing chain length. In addition, we have shown that for heavier polyatomic molecules it can be advantageous to attach two OCCs, thereby potentially doubling the photon scattering rate and thus speeding up cooling rates. 

We also demonstrated that the OCC can be attached to aromatic molecules, such as the C$_{60}$ fullerene, by a rigid C$\equiv$C link that prevents  strong 
interaction with the OCC by keeping the metal center away from the delocalized $\pi$ electrons of the aromatic molecule. This kind of link also 
retains the cycling properties of alkaline-earth atoms in the endohedral N@C$_{60}$ complex. This type of complexes is of great interest as building blocks for a scalable fullerene-based high-spin quantum registers of quantum computer~\cite{Benjamin2006}.  

We finish by mentioning some open questions that future research could answer: what other types of metallic atom and chains can be used? What  chemical bond is necessary to retain the beneficial photon scattering properties seen in polyatomic molecules?  Improving our understanding of the photon cycling properties of the simplest molecules will also elucidate OCCs and point toward the most promising paths toward increasing molecular complexity.

\begin{acknowledgments}
Work at  Temple University is supported by the Army Research Office
Grant No. W911NF-17-1-0563, the U.S. Air Force Office of Scientific 
Research Grant No. FA9550-14-1-0321 and the NSF Grant No. PHY-1908634. 
\end{acknowledgments}

\pagebreak

\bibliography{OCC_Library}

\end{document}